\documentclass[aps,prd,preprintnumbers,groupedaddress,showpacs,superscriptaddress,manuscript,floatfix]{revtex4}
\usepackage{epsfig}
\usepackage{latexsym}
\usepackage{graphicx}
\usepackage{bm}
\usepackage{longtable}
\usepackage{color}
\usepackage{ulem}
\usepackage{amssymb,amsmath}

\newcommand{\kslash}{k\kern-1ex /}
\newcommand{\pslash}{p\kern-1ex /}
\newcommand{\qslash}{q\kern-1ex /}
\newcommand{\lslash}{l\kern-1ex /}
\newcommand{\sslash}{s\kern-1ex /}
\newcommand{\Dslash}{D\kern-1.2ex /}

\newcommand{\beqa}{\begin{eqnarray}}
\newcommand{\eeqa}{\end{eqnarray}}

\newcommand{\bd}{\begin{description}}
\newcommand{\ed}{\end{description}}

\newcommand{\ben}{\begin{eqnarray}}
\newcommand{\een}{\end{eqnarray}}

\def\lsim{\raise0.3ex\hbox{$<$\kern-0.75em\raise-1.1ex\hbox{$\sim$}}}
\def\gsim{\raise0.3ex\hbox{$>$\kern-0.75em\raise-1.1ex\hbox{$\sim$}}}
\def\simgt{\rlap{\lower 3.5 pt\hbox{$\mathchar \sim$}}\raise 1.0pt \hbox {$>$}}
\def\simlt{\rlap{\lower 3.5 pt\hbox{$\mathchar \sim$}}\raise 1.0pt \hbox {$<$}}

\begin{document}

\bibliographystyle{apsrev}

\preprint{UTCCS-P-67, UTHEP-642}

\title{Helium nuclei, deuteron and dineutron in 2+1 flavor lattice QCD}

\author{Takeshi Yamazaki}
\affiliation{Kobayashi-Maskawa Institute for the Origin of Particles and the Universe, Nagoya University, Naogya, Aichi 464-8602, Japan}
\affiliation{RIKEN Advanced Institute for Computational Science,
Kobe, Hyogo 650-0047, Japan}
\author{Ken-ichi Ishikawa}
\affiliation{Department of Physics,
Hiroshima University,
Higashi-Hiroshima, Hiroshima 739-8526, Japan}
\affiliation{RIKEN Advanced Institute for Computational Science,
Kobe, Hyogo 650-0047, Japan}
\author{Yoshinobu Kuramashi}
\affiliation{Center for Computational Sciences, University of Tsukuba,
Tsukuba, Ibaraki 305-8577, Japan}
\affiliation{Graduate School of Pure and Applied Sciences,
University of Tsukuba, Tsukuba, Ibaraki 305-8571, Japan}
\affiliation{RIKEN Advanced Institute for Computational Science,
Kobe, Hyogo 650-0047, Japan}
\author{Akira Ukawa}
\affiliation{Center for Computational Sciences, University of Tsukuba,
Tsukuba, Ibaraki 305-8577, Japan}

\date{
\today
}

\begin{abstract}

We calculate the binding energies for multi-nucleon bound states with
the nuclear mass number less than or equal to 4 
in 2+1 flavor QCD at the lattice spacing of $a = 0.09$ fm
employing a relatively heavy quark mass corresponding to $m_\pi = 0.51$ GeV.
To distinguish a bound state from attractive scattering states,
we investigate the volume dependence of the energy shift 
between the ground state and the state of free nucleons by changing the
spatial extent of the lattice from 2.9 fm to 5.8 fm.
We conclude that $^4$He, $^3$He,
deuteron and dineutron are bound at $m_\pi = 0.51$ GeV.
We compare their binding energies with those in our quenched studies
and also with several previous investigations.

\end{abstract}

\maketitle

\section{Introduction}
\label{sec:introduction}

Lattice QCD has a potential ability to quantitative understand the nature of nuclei,
whose characteristic feature is a hierarchical structure in the strong interaction. 
The nuclear binding energy is experimentally known to be
about 10 MeV per nucleon,  which is much smaller than the typical energy scale of hadrons.
A measurement of the binding energies is therefore 
the first step for direct investigation of nuclei in lattice QCD.  
A key ingredient in the study is a systematic change of the spatial volume of the lattice to
distinguish a bound state from an attractive scattering state.

We carried out a first attempt to measure   
the binding energies of the $^4$He and $^3$He nuclei in quenched QCD 
with a rather heavy quark mass corresponding to $m_\pi=0.80$ GeV, thereby  avoiding heavy computational cost~\cite{Yamazaki:2009ua}. 
We followed this work by a renewed investigation of the bound state for the two-nucleon channel 
in quenched QCD at the same quark mass, which found that not only the deuteron 
in the $^3$S$_1$ channel 
but also the dineutron in the $^1$S$_0$ channel is bound~\cite{Yamazaki:2011nd}.
Independently, NPLQCD collaboration reported 
the possibility that a bound state is formed in both channels at $m_\pi=0.39$ GeV in 2+1 flavor QCD~\cite{Beane:2011iw}.  They later confirmed the bound states for the helium nuclei and the two-nucleon channels at $m_\pi=0.81$ GeV in 3-flavor QCD taking a different choice for the quark and gluon actions~\cite{Beane:2012vq}. 

In this article we report on our investigation of the dynamical quark effects on the binding energies of the helium nuclei, 
the deuteron and the dineutron.  We perform 2+1 flavor lattice QCD simulation with the degenerate 
up and down quark mass corresponding to $m_\pi=0.51$ GeV. Four lattice sizes are employed to take the 
infinite spatial volume limit: $32^3\times 48$, $40^3\times 48$, $48^3\times 48$and $64^3\times 64$, whose spatial extent 
ranges from 2.9 fm to 5.8 fm with the lattice spacing of $a=0.08995(40)$ fm~\cite{Aoki:2009ix}. 

For the helium nuclei our main interest lies 
in the magnitude of the binding energies, since all studies carried out so far, both  in quenched and in unquenched QCD and for several quark mass values,  agree on the bound state nature for helium nuclei.  
Much more intriguing is the  two-nucleon system, for which there are two ways to study. 
One is a direct investigation~\cite{Fukugita:1994na,Fukugita:1994ve,Beane:2006mx,Beane:2009py,Yamazaki:2011nd,Beane:2011iw,Beane:2012vq} in which one calculates the two-nucleon Green's functions directly in lattice QCD, and the other is 
an  indirect calculation by means of the two-nucleon effective potential extracted from 
the two-nucleon wave function in lattice QCD~\cite{Ishii:2006ec,Aoki:2009ji}.

So far only the former method has reported the binding energies of the
two-nucleon systems. In quenched QCD the bound state nature has been confirmed 
for both channels at $m_\pi=0.80$~GeV in our recent work~\cite{Yamazaki:2011nd}. 
On the other hand, unquenched studies show a complicated situation.  A somewhat early 
study in 2+1 flavor QCD with a mixed action~\cite{Beane:2006mx} reported a positive energy shift (repulsive interaction) 
in both channels at $m_\pi\le 0.59$ GeV.  More recently, however, deep bound states were  observed at $m_\pi=0.81$ GeV in 3-flavor QCD~\cite{Beane:2012vq}.  
We hope to shed light on this situation with our own investigation in 2+1 flavor QCD.

This paper is organized as follows. In Sec.~\ref{sec:details} we explain simulation details including 
the simulation parameters and the interpolating operators for the multi-nucleon channels. Section~\ref{sec:results} presents the results 
of the binding energies for the helium nuclei, the deuteron and the dineutron. We compare our results with those 
in the previous studies. Conclusions and discussions are
summarized in Sec.~\ref{sec:summary}.

\section{Simulation details}
\label{sec:details}
\subsection{Simulation parameters}
\label{sec:sim_para}

We generate 2+1 flavor gauge configurations with the 
Iwasaki gauge action~\cite{Iwasaki:2011jk} and 
the non-perturbative $O(a)$-improved Wilson quark action
at $\beta = 1.90$ with $c_{\rm SW} = 1.715$~\cite{Aoki:2005et}.
The lattice spacing is $a=0.8995(40)$ fm, corresponding to $a^{-1} = 2.194(10)$ GeV,
determined with $m_\Omega=1.6725$ GeV~\cite{Aoki:2009ix}.
We take four lattice sizes, 
$L^3\times T = 32^3 \times 48$, $40^3 \times 48$, $48^3 \times 48$ 
and $64^3 \times 64$, 
to investigate the spatial volume dependence of the ground state energy 
shift between the multi-nucleon system and the free nucleons.
The physical spatial extents are 2.9, 3.6, 4.3 and 5.8 fm, respectively.
Since it becomes harder to obtain a good signal-to-noise ratio at lighter quark masses for multi-nucleon systems~\cite{Lepage:1989hd,Fukugita:1994ve},
we employ the hopping parameters $(\kappa_{ud},\kappa_s) = (0.1373316,0.1367526)$ 
which correspond to $m_\pi = 0.51$ GeV and $m_N = 1.32$ GeV and the physical value for the strange quark mass. These values are chosen based on the previous results for $m_\pi$ and
$m_s$ obtained by the PACS-CS Collaboration~\cite{Aoki:2008sm,Aoki:2009ix}.

We employ the domain-decomposed Hybrid-Monte-Carlo (DDHMC) 
algorithm~\cite{Luscher:2003vf,Luscher:2005rx}
for the degenerate light quarks and the UV-filtered PHMC (UVPHMC)
algorithm~\cite{Ishikawa:2006pb} for the strange quark
 employing the 
Omelyan-Mryglod-Folk integrator~\cite{{Omelyan:2003om},{Takaishi:2005tz}}.
The algorithmic details are given in Ref.~\cite{Aoki:2008sm}.
We summarize simulation parameters in Table~\ref{tab:conf_gene} including
the block sizes in DDHMC and the polynomial order in UVPHMC.
We take $\tau = 1$ for the trajectory length of the molecular dynamics in all the runs. The step sizes are
chosen such that we obtain reasonable acceptance rates presented 
in Table~\ref{tab:conf_gene}.
We generate the gauge configurations in a single run except
for the $L=64$ case for which we carry out two runs.
The total trajectory length is about 2000 for all the volumes, 
except 4000 for  the case of smallest volume.

\subsection{Calculation method}
\label{sec:method}

We extract the ground state energies of the multi-nucleon systems 
and the nucleon state from the correlation functions
\begin{equation}
G_{\mathcal{O}}(t) = \langle 0 | \mathcal{O}(t)
\overline{\mathcal{O}}(0) | 0 \rangle
\label{eq:corr}
\end{equation}
with $\mathcal{O}$ being appropriate operators
for $^4$He, $^3$He, two-nucleon $^3$S$_1$ and $^1$S$_0$ channels,  
and the nucleon state $N$ (see the next subsection for actual expressions). 

We carry out successive measurements in the interval of 10 trajectories.
The errors are estimated by jackknife analysis choosing 200
trajectories for the bin size for all volumes,
except for the largest volume for which we use 190.
The numbers of configurations are listed in Table~\ref{tab:conf_meas}.
We attempt to extract as much information as possible from each configuration 
by repeating the measurement of the correlation functions for a number of sources at different spatial points and time slices.
For the $48^4$ and $64^4$ lattices, we calculate the correlation functions not only in the temporal direction but also in the three spatial directions exploiting the space-time rotational symmetry.
We found that this procedure effectively increases statistics by a factor of four. 
This factor is included in
the number of measurements on each configuration
given in Table~\ref{tab:conf_meas}

We are interested in the energy shift between the ground state of the multi-nucleon system and the free nucleons on an $L^3$ box,
\begin{equation}
\Delta E_L=E_{\mathcal{O}}-N_N m_N
\label{eq:delE_L}
\end{equation}
with $E_{\mathcal{O}}$ being the lowest energy level for the multi-nucleon
channel, $N_N$ the number of nucleon and $m_N$ the nucleon mass. 
This quantity is directly extracted from the ratio of the multi-nucleon 
correlation function
divided by the $N_N$-th power of the nucleon correlation function
\begin{equation}
R(t) = \frac{G_{\mathcal{O}}(t)}{\left(G_N(t)\right)^{N_N}},
\label{eq:R}
\end{equation}
where the same source operator is chosen for the numerator
and the denominator.
We also define the effective energy shift as
\begin{equation}
\Delta E_L^{\mathrm{eff}} = \ln \left(\frac{R(t)}{R(t+1)}\right),
\label{eq:eff_delE_L}
\end{equation}
which is useful to check the plateau region in later sections.
  
Note that the definitions for $\Delta E_L$ and $\Delta E_L^{\mathrm{eff}}$ 
follow those in Ref.~\cite{Yamazaki:2011nd}, but are opposite to those used in 
Ref.~\cite{Yamazaki:2009ua}.

\subsection{Interpolating operators}
\label{sec:operator}

We use an interpolating operator for the proton given by 
\begin{equation}
p_\alpha = \varepsilon_{abc}([u_a]^tC\gamma_5 d_b)u_c^\alpha,
\label{eq:def:proton}
\end{equation}
where $C = \gamma_4 \gamma_2$ and $\alpha$ and $a,b,c$ are the Dirac index and
the color indices, respectively.
The neutron operator $n_\alpha$ is obtained 
by replacing $u_c^\alpha$ by $d_c^\alpha$ 
in the proton operator.
To save the computational cost
we use the non-relativistic quark operator, in which the Dirac index
is restricted to the upper two components.

The $^4$He nucleus has zero total angular momentum, positive parity
$J^P = 0^+$ and  zero isospin $I = 0$. 
We employ the simplest $^4$He interpolating operator with 
zero orbital angular momentum $L=0$, and hence $J=S$ with $S$ being
the total spin.
Such an operator was already given long time ago in Ref.~\cite{Beam:1967zz},
\begin{equation}
^4\mathrm{He} = 
\frac{1}{\sqrt{2}}\left( \overline{\chi}\eta - 
\chi \overline{\eta} \right),
\end{equation}
where 
\begin{eqnarray}
\chi &=&  \frac{1}{2}( [+-+-] + [-+-+] - [+--+] - [-++-] ),\\
\overline{\chi}  &=& \frac{1}{\sqrt{12}}( 
[+-+-] + [-+-+] + [+--+] + [-++-] - 2 [++--] - 2 [--++] 
)
\end{eqnarray}
with $+/-$ being up/down spin of each nucleon, and  
$\eta, \overline{\eta}$ are obtained
by replacing $+/-$ in $\chi, \overline{\chi}$ by $p/n$ for the isospin.
Each nucleon in the sink operator is projected to zero spatial
momentum.

We also calculate the correlation function of 
the $^3$He nucleus whose quantum numbers are 
$J^P=\frac{1}{2}^+$, $I = \frac{1}{2}$ and $I_z = \frac{1}{2}$.
We employ the interpolating operator in Ref.~\cite{Bolsterli:1964zz},
\begin{equation}
^3{\rm He} = 
\frac{1}{\sqrt{6}}\left(
\left|p_- n_+ p_+ \right\rangle
-
\left|p_+ n_+ p_- \right\rangle
+
\left|n_+ p_+ p_- \right\rangle
-
\left|n_+ p_- p_+ \right\rangle
+
\left|p_+ p_- n_+ \right\rangle
-
\left|p_- p_+ n_+ \right\rangle
\right),
\end{equation}
with the zero momentum projection on each nucleon in the sink operator.

The two-nucleon operators for the $^3$S$_1$ and $^1$S$_0$ 
channels are given by 
\begin{eqnarray}
NN_{^3{\mathrm S}_1}(t) &=& \frac{1}{\sqrt{2}}\left[
p_+(t) n_+(t) - n_+(t) p_+(t)
\right],
\label{eq:def:3S1}\\
NN_{^1{\mathrm S}_0}(t) &=& 
\frac{1}{\sqrt{2}}\left[
n_+(t) n_-(t) - n_-(t) n_+(t)
\right].
\label{eq:def:1S0}
\end{eqnarray}
In the spin triplet channel
the operators for the other two spin components are constructed in a similar way.
We take average over the three spin components.

The quark propagators are solved with
the periodic boundary condition 
in all the spatial and temporal directions
using the exponentially smeared source of form 
\begin{equation}
q^\prime(\vec{x},t) = \sum_{\vec{y}} A\, e^{-B|{\vec x} - \vec{y}|} q(\vec{y},t)
\label{eq:smear}
\end{equation}
after the Coulomb gauge fixing.
We choose the smearing parameters depending on the volume (see Table~\ref{tab:conf_meas}) 
in order to obtain reasonable plateaux of the effective energy for 
the ground states in the multi-nucleon channels as well as for the nucleon.
For the source operators explained above we insert the {\it smeared} quark fields of 
Eq.~(\ref{eq:smear}) for each nucleon operator located at the same spatial point $\vec{x}$.
Each nucleon in the sink operator, on the other hand, 
is composed of the {\it point} quark fields,  
and projected to zero spatial
momentum.

\subsection{Difficulties for multi-nucleon channel}
\label{sec:problem}

There are several computational difficulties in
the calculation of the correlation functions $G_{\mathcal{O}}(t)$
for the $^3$He and $^4$He channels.
One is a factorially large number of 
Wick contractions for the quark-antiquark fields.
A naive counting gives $(2N_p + N_n)!(2N_n + N_p)!$ for a nucleus 
composed of $N_p$ protons and $N_n$ neutrons, which quickly becomes 
prohibitively large beyond three-nucleon systems, {\it e.g., } 2,880 for 
$^3$He and 518,400 for $^4$He.  
To overcome the difficulty, we use the reduction techniques proposed in 
our exploratory work~\cite{Yamazaki:2009ua}.
After the reduction, only 1107 (93) 
contractions are required for the correlation function
in the $^4$He ($^3$He) channel.
Other reduction techniques for the large number of the Wick contractions 
have been proposed for the multi-meson~\cite{Detmold:2010au}
and multi-baryon~\cite{{Doi:2012xd},{Detmold:2012aa}} channels.

Another difficulty in studying a multi-nucleon bound state is
the identification of the bound state nature in a finite volume,
because an attractive scattering state yields a  similar
energy shift due to the finite volume effect~\cite{{Luscher:1986pf},{Beane:2003da},{Sasaki:2006jn}}.
To solve the problem we need to investigate the volume dependence
of the measured energy shift~\cite{Yamazaki:2009ua,Yamazaki:2011nd}:
For a scattering state, the energy shift decreases in proportion to $1/L^3$
at the leading order in the $1/L$ expansion~\cite{Luscher:1986pf,Beane:2007qr}, 
while for a bound state the energy shift remains at a finite value 
in the infinite spatial volume limit.
In order to distinguish
a non-zero constant from a $1/L^3$ behavior in the energy shift,
we employ four spatial extents from 2.9 to 5.8 fm.

\section{Results}
\label{sec:results}

\subsection{Nucleon} 

\begin{figure}[!t]
\includegraphics*[angle=0,width=0.6\textwidth]{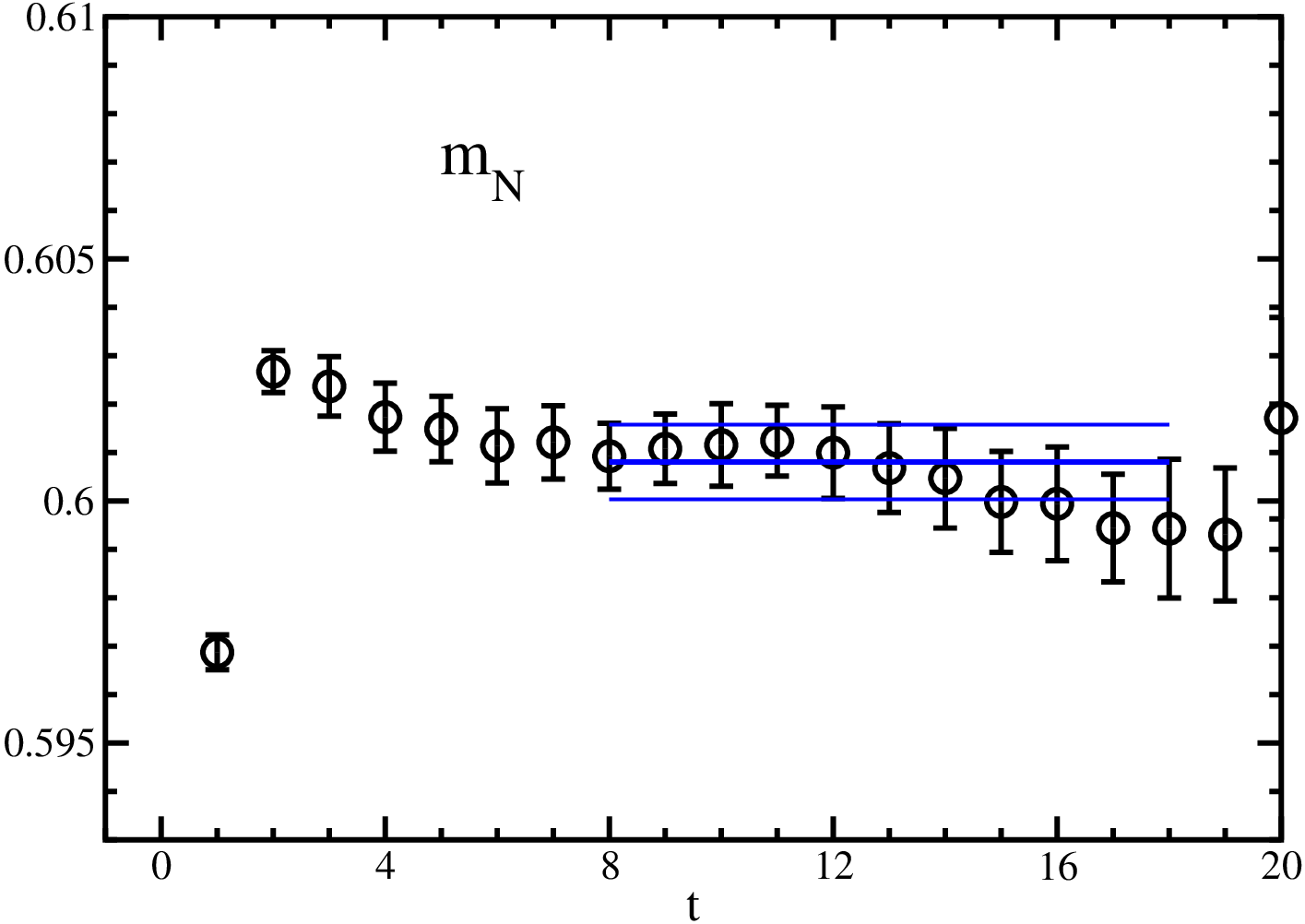}
\caption{
Nucleon effective mass on 
(5.8 fm)$^3$ box in lattice unites.
Fit result with one standard deviation error band is expressed
by solid lines.
\label{fig:eff_N}
}
\end{figure}

We first show the effective nucleon mass on the (5.8 fm)$^3$ box 
in Fig.~\ref{fig:eff_N} as a typical result.
The plateau of the effective mass is clearly observed. 
A fit result of the correlation
function with an exponential form is also drawn in the figure 
with the one standard deviation error band.
We list the nucleon mass together with the pion mass in
Table~\ref{tab:conf_meas}.

\subsection{$^4$He nucleus}
\label{sec:4N}

\begin{figure}[!t]
\includegraphics*[angle=0,width=0.6\textwidth]{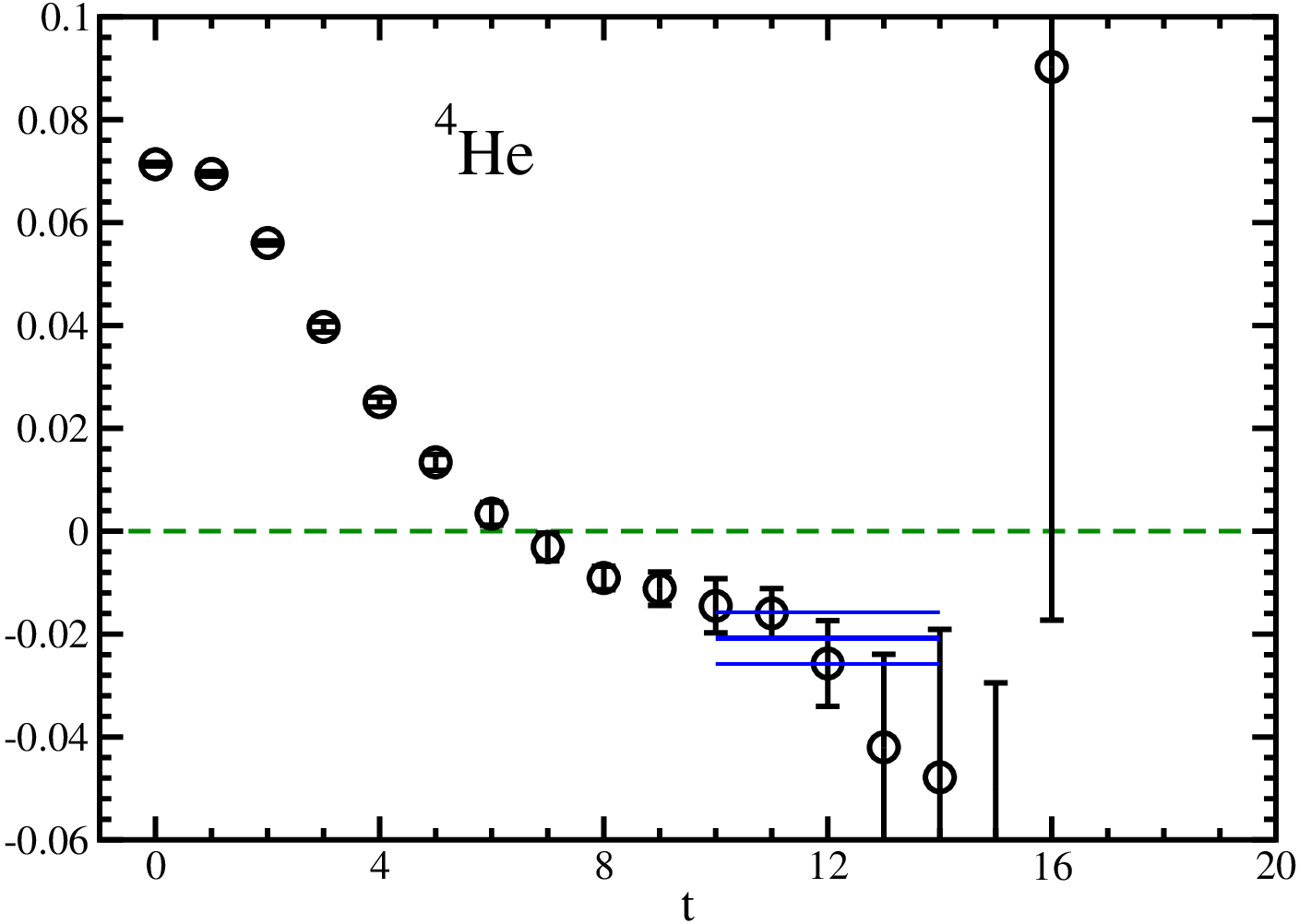}
\caption{
Effective energy shift 
$\Delta E_L^{\mathrm{eff}}$
for $^4$He channel on (5.8 fm)$^3$ box in lattice units.
Fit result with one standard deviation error band is expressed
by solid lines.
\label{fig:eff_R_h4}
}
\end{figure}

The effective energy shift $\Delta E^{\mathrm{eff}}_L$ defined in
Eq.~(\ref{eq:eff_delE_L}) is plotted in Fig.~\ref{fig:eff_R_h4}.
The signal is clear up to $t=12$, 
beyond which the statistical error increases rapidly.
The energy shift $\Delta E_L$ is extracted from 
$R(t)$ of Eq.~(\ref{eq:R}) by an exponential fit
over the range of $t = 10$--14.
The fit result is denoted by the solid lines with the statistical error
band in Fig.~\ref{fig:eff_R_h4}.
The systematic error in the fit 
is estimated from the variation of the fit results with
the minimum or maximum time slice changed by $\pm 1$.
Results with similar quality are obtained on other volumes.
We summarize the values of $\Delta E_L$ with the statistical 
and systematic errors in Table~\ref{tab:dE_He}.

\begin{figure}[!t]
\includegraphics*[angle=0,width=0.6\textwidth]{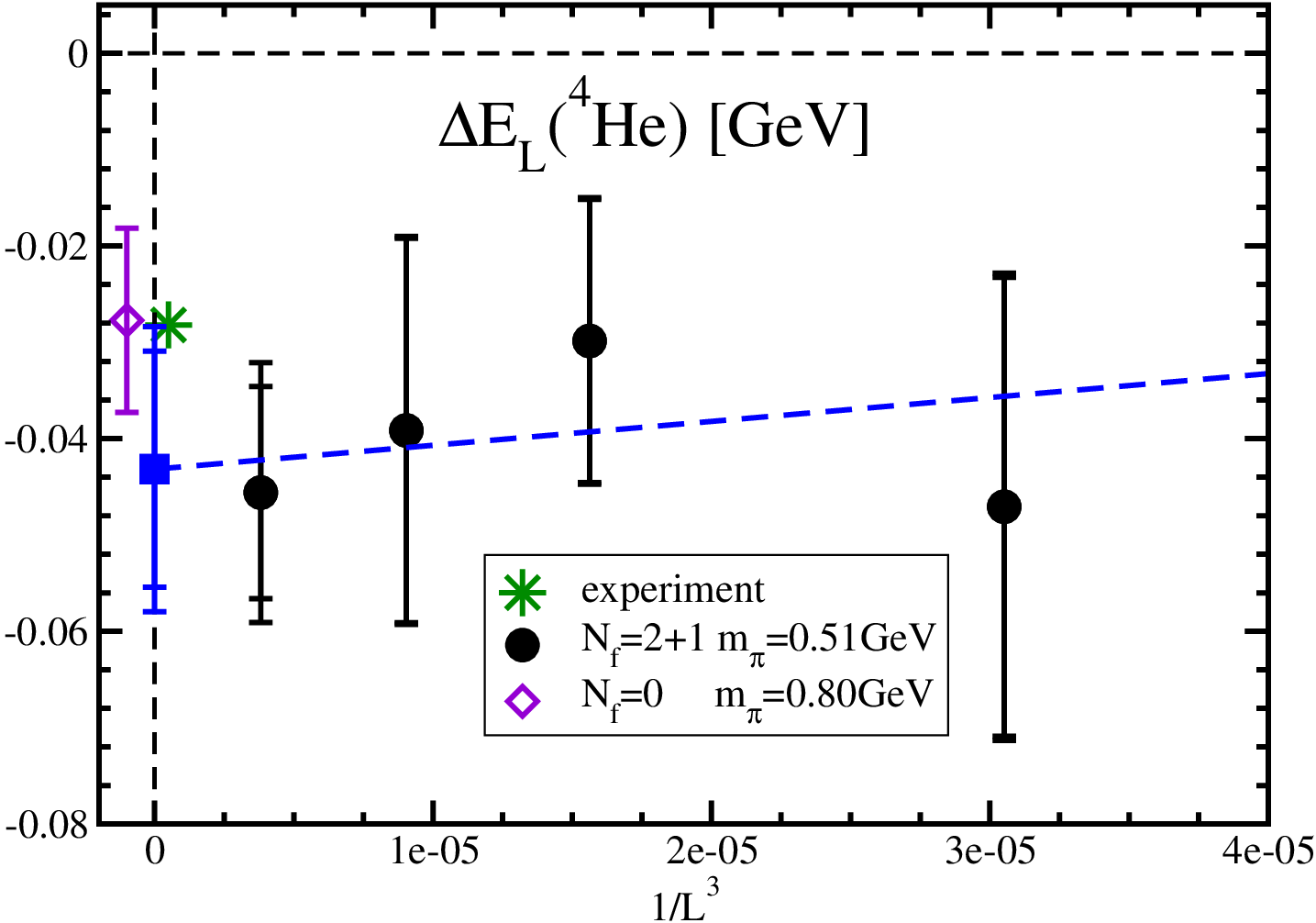}
\caption{
Spatial volume dependence of $\Delta E_L$ in GeV units
for $^4$He channel.
Outer bar denotes the combined error of statistical and 
systematic ones added in quadrature. Inner bar is for
the statistical error.
Extrapolated result in the infinite spatial volume limit is
shown by filled square symbol together with the fit line (dashed).
Experimental value (star) and quenched result (open diamond) 
are also presented.
\label{fig:dE_h4}
}
\end{figure}

Figure~\ref{fig:dE_h4} shows the volume dependence of $\Delta E_L$
as a function of $1/L^3$.
The inner bar denotes the statistical error and the outer bar represents the statistical and 
systematic errors combined in quadrature.
The negative energy shifts are obtained in all the four volumes.
We extrapolate the results
to the infinite volume limit with a simple 
linear function of $1/L^3$,
\begin{equation}
\Delta E_L = \Delta E_\infty + \frac{C_L}{L^3}.
\label{eq:linear}
\end{equation}
The systematic error is estimated from the variation of
the results obtained by alternative
fits which contains a constant fit of the data 
and a fit of the data obtained with a different fit range in $t$.
The non-zero negative value obtained for the infinite volume limit $\Delta E_\infty$ 
shown in Fig.~\ref{fig:dE_h4} and Table~\ref{tab:dE_He}
leads us to conclude  that the ground state is bound in this channel for the quark masses employed. 
The binding energy $-\Delta E_\infty=43(12)(8)$ MeV, where the first error is statistical and the second one is systematic, 
is consistent with the experimental result of 28.3 MeV and 
also with the previous quenched result at
$m_\pi = 0.80$ GeV~\cite{Yamazaki:2011nd}.  Note that the error is still quite large. 

A recent work in 3-flavor QCD at $m_\pi=0.81$ GeV reported a value 
110(20)(15) MeV for the binding energy of $^4$He nucleus~\cite{Beane:2012vq}.  
This is about three times deeper than our value.  Whether this difference can be attributed to 
the quark mass dependence in unquenched calculations
needs to be clarified in future. 

\subsection{$^3$He nucleus}
\label{sec:3N}

\begin{figure}[!t]
\includegraphics*[angle=0,width=0.6\textwidth]{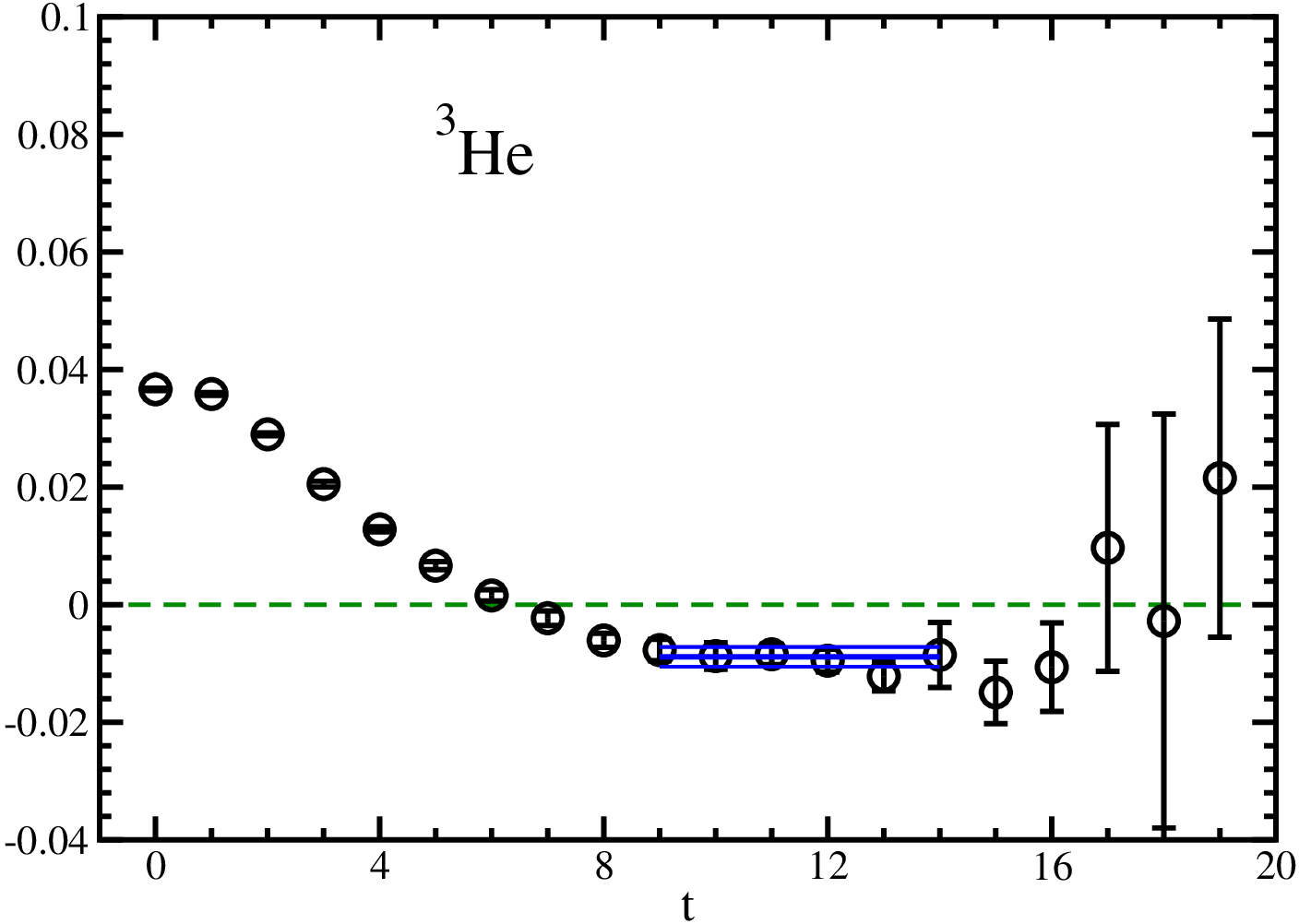}
\caption{
Same as Fig.~\ref{fig:eff_R_h4} for $^3$He channel.
\label{fig:eff_R_h3}
}
\end{figure}

Figure \ref{fig:eff_R_h3} shows the effective energy shift 
$\Delta E^{\mathrm{eff}}_L$ of Eq.~(\ref{eq:eff_delE_L}).
The quality of the signal is better than the 
$^4$He channel in Fig.~\ref{fig:eff_R_h4}.
An exponential fit of $R(t)$ in Eq.~(\ref{eq:R})
with the range of $t = 9$--14 yields
a negative value, which is denoted by the solid lines 
with the statistical error band in Fig.~\ref{fig:eff_R_h3}.
The systematic error in the fit 
is estimated in the same way as in the $^4$He case.

\begin{figure}[!t]
\includegraphics*[angle=0,width=0.6\textwidth]{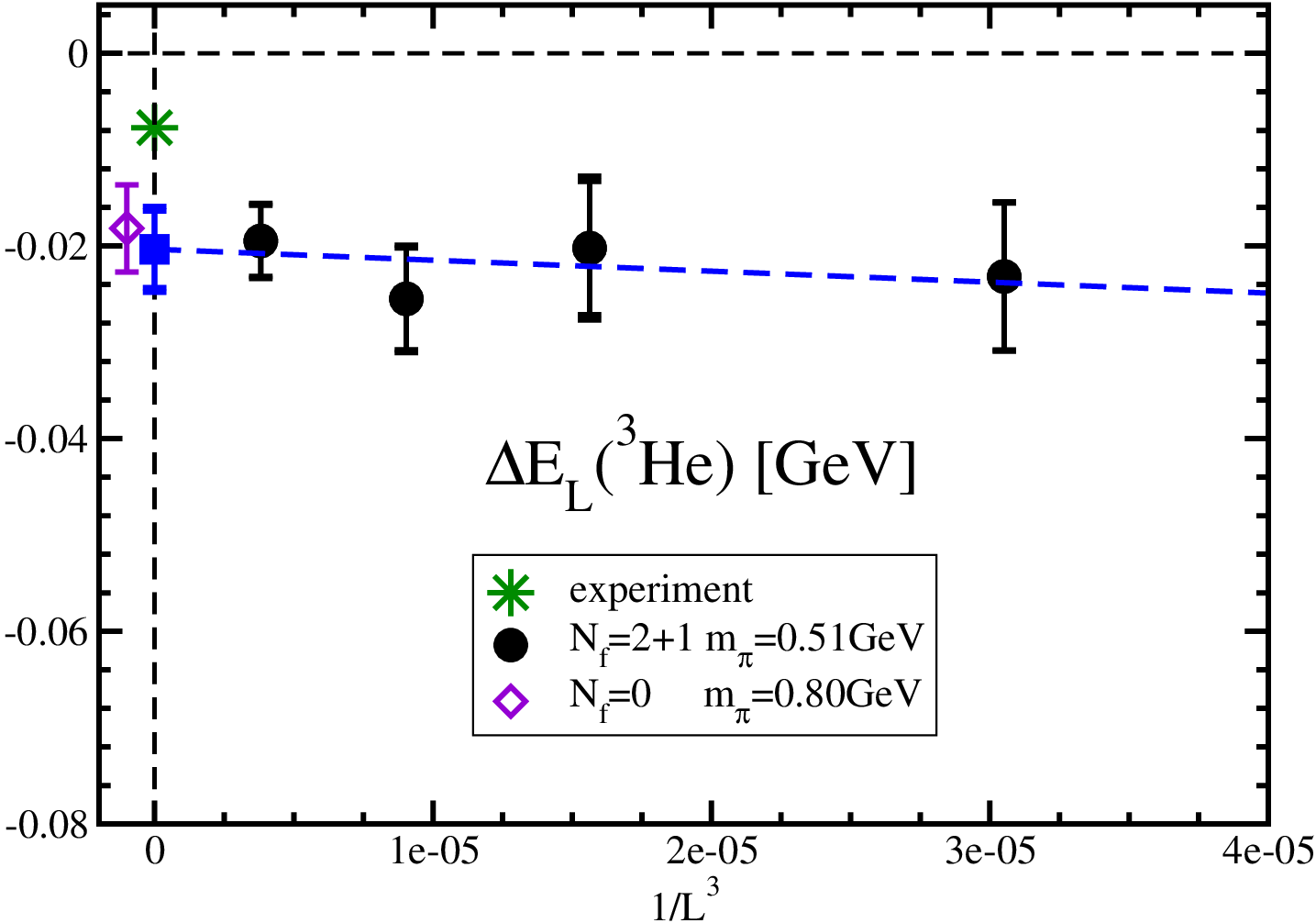}
\caption{
\label{fig:dE_h3}
Same as Fig.~\ref{fig:dE_h4} for $^3$He channel.
}
\end{figure}

As listed in Table~\ref{tab:dE_He} we find non-zero negative values 
for the energy shift $\Delta E_L$ for all the volumes.
The volume dependence is illustrated in Fig.~\ref{fig:dE_h3} as a function
of $1/L^3$ with the inner and outer error bars 
as explained in the previous subsection. 
We carry out a linear extrapolation of Eq.~(\ref{eq:linear}).
The systematic error is estimated in the same way as in the $^4$He channel.
The energy shift extrapolated to the infinite spatial volume limit is non-zero and negative, 
see Fig.~\ref{fig:dE_h3}  and Table~\ref{tab:dE_He},
which means that the ground state is a bound state in this channel.
The value of $-\Delta E_\infty=20.3(4.0)(2.0)$ MeV is 
roughly three times larger than
the experimental result, 7.72 MeV,
though consistent with our previous quenched result at
$m_\pi = 0.80$ GeV~\cite{Yamazaki:2011nd}.

In 3-flavor QCD $-\Delta E_\infty=71(6)(5)$ MeV was reported~\cite{Beane:2012vq} 
at a heavier quark mass corresponding to $m_\pi=0.81$ GeV.  
Here again future work is needed to see if a quark mass dependence explains the difference from the experiment.

\subsection{Two-nucleon channels}
\label{sec:2N}
\subsubsection{Present work}

\begin{figure}[!t]
\includegraphics*[angle=0,width=0.6\textwidth]{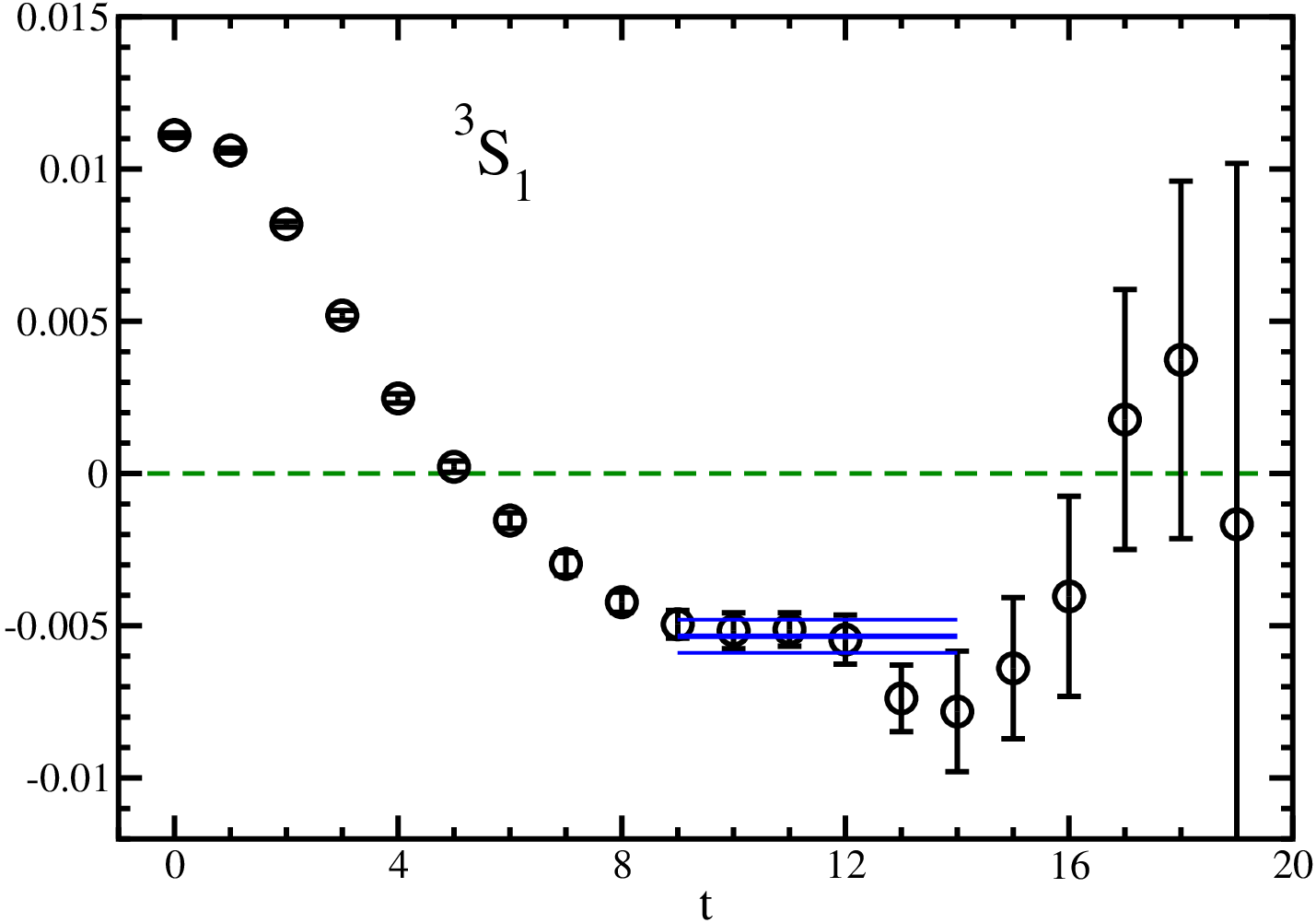}
\caption{
Same as Fig.~\ref{fig:eff_R_h4} for $^3$S$_1$ channel.
\label{fig:eff_R_3S1}
}
\end{figure}

In Fig.~\ref{fig:eff_R_3S1} we show the 
time dependence for $\Delta E_L^{\mathrm{eff}}$ of Eq.~(\ref{eq:eff_delE_L})
in the $^3$S$_1$ channel.
The signals 
are lost beyond $t\approx 14$.
We observe negative values
beyond the error bars in the plateau region of $t=9$--14.
We extract the value of $\Delta E_L$ from an exponential fit for $R(t)$ of
Eq.~(\ref{eq:R}) in the range of $t=9$--14.
The systematic error of the fit 
is estimated as explained in the previous subsections.

\begin{figure}[!t]
\includegraphics*[angle=0,width=0.6\textwidth]{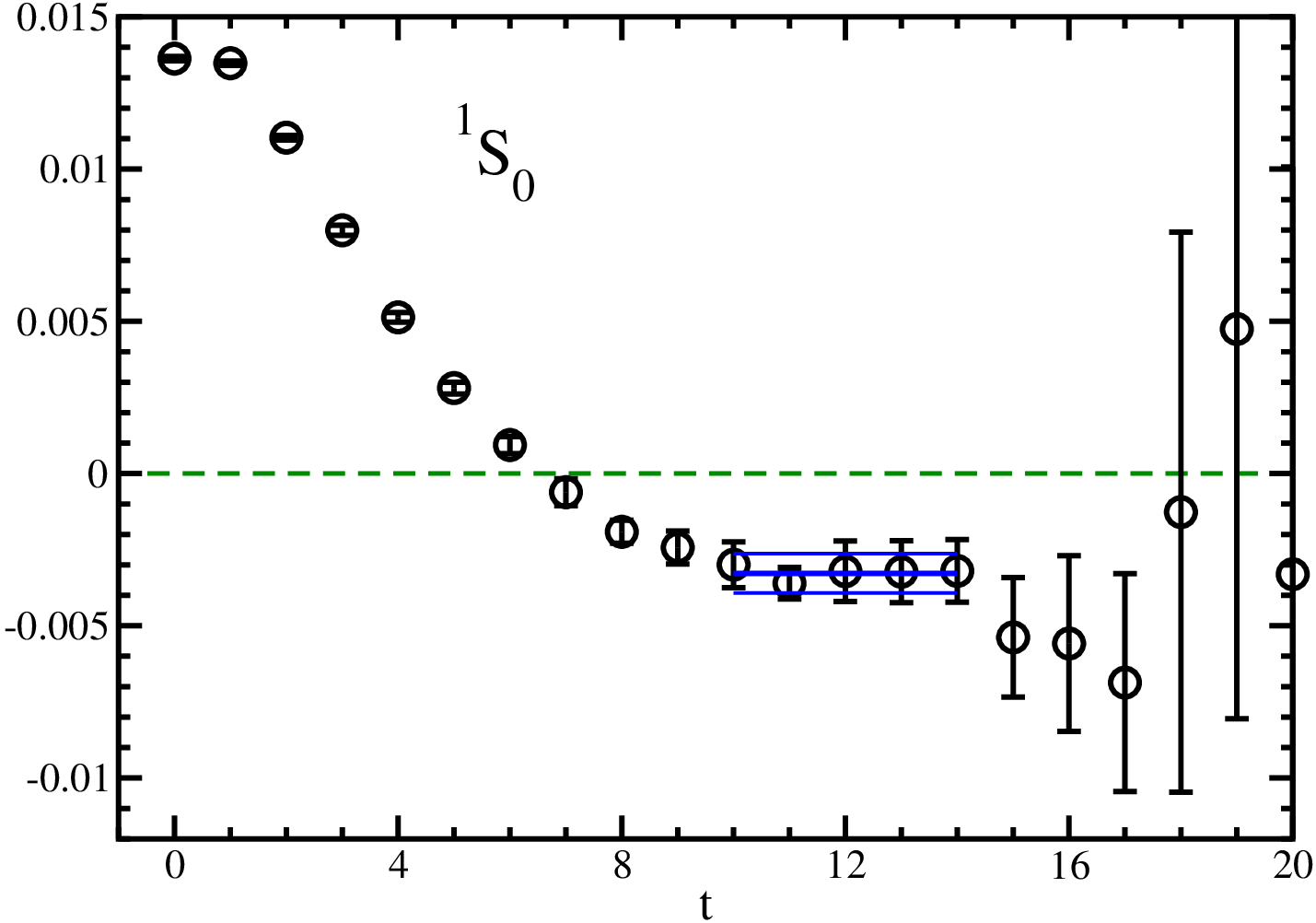}
\caption{
Same as Fig.~\ref{fig:eff_R_h4} for $^1$S$_0$ channel.
\label{fig:eff_R_1S0}
}
\end{figure}

Figure~\ref{fig:eff_R_1S0} shows the result for $\Delta E_L^{\mathrm{eff}}$ 
in the $^1$S$_0$ channel on the (5.8 fm)$^3$ box.
The value of $\Delta E_L^{\mathrm{eff}}$ 
is again negative beyond the error bars in the plateau region,
though the absolute value is smaller than in the $^3$S$_1$ case.
The energy shift $\Delta E_L$ is obtained in the same way 
as for the $^3$S$_1$ channel.

\begin{figure}[!t]
\includegraphics*[angle=0,width=0.6\textwidth]{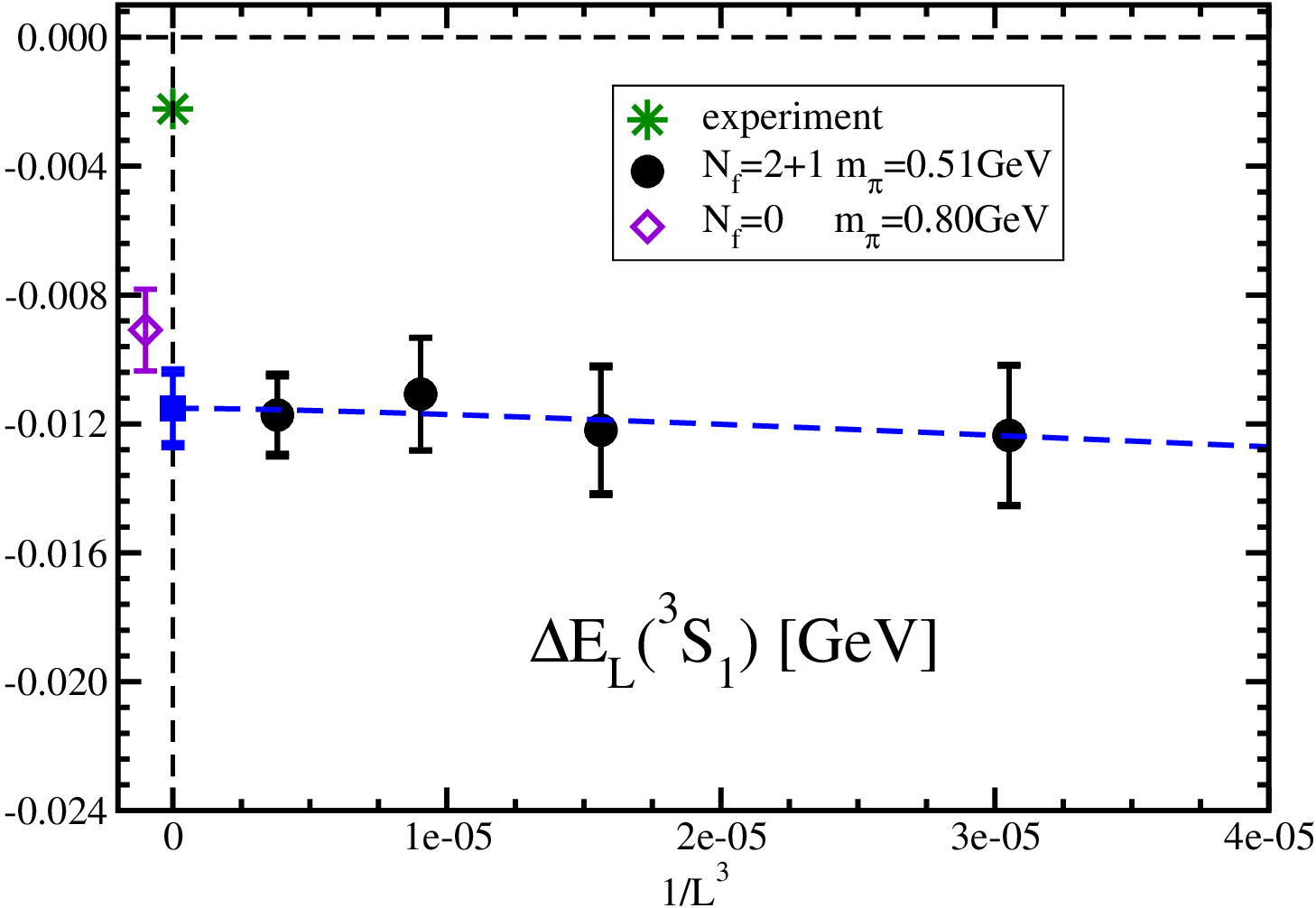}
\caption{
Same as Fig.~\ref{fig:dE_h4} for $^3$S$_1$ channel.
\label{fig:dE_3S1}
}
\end{figure}

\begin{figure}[!t]
\includegraphics*[angle=0,width=0.6\textwidth]{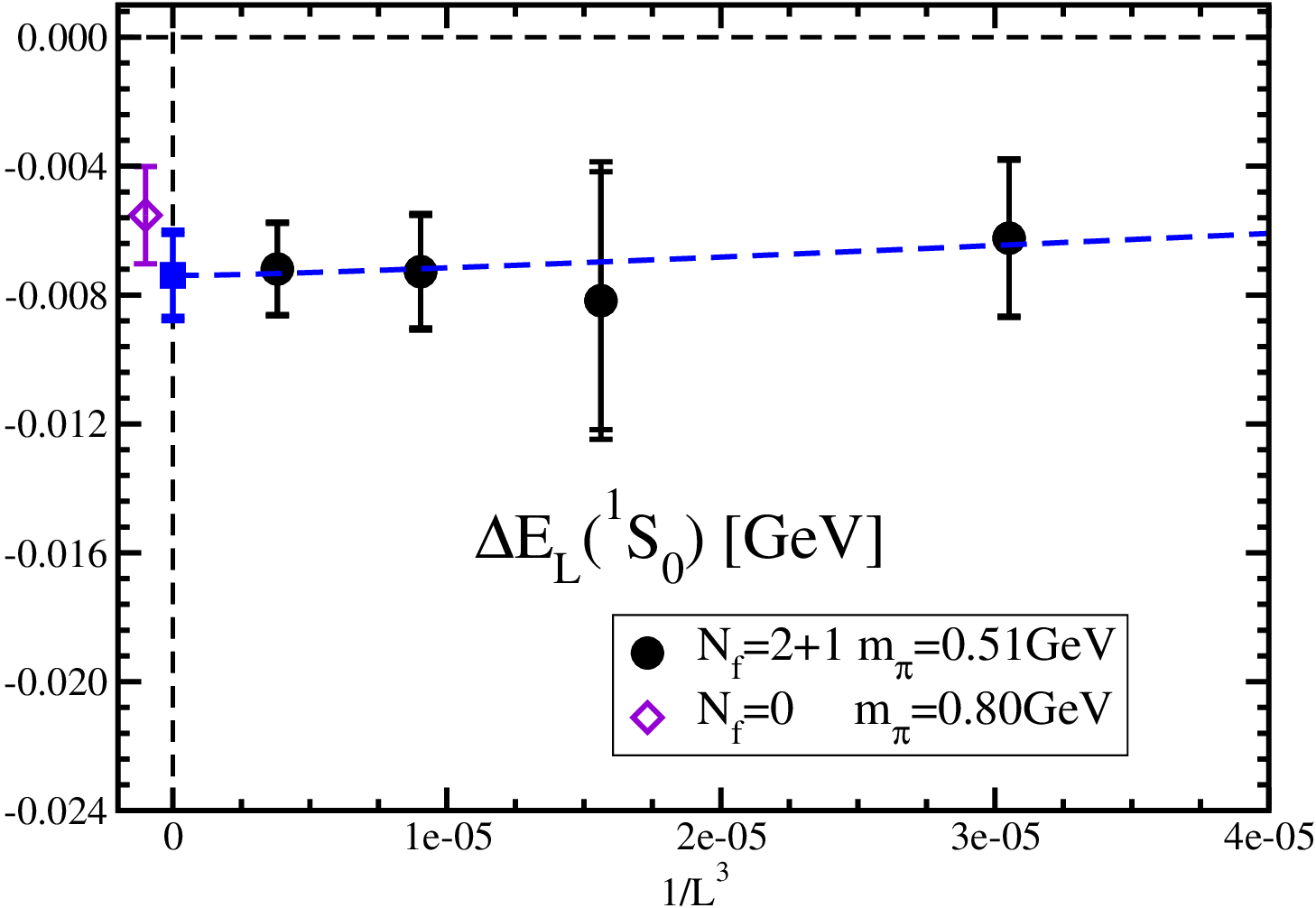}
\caption{
Same as Fig.~\ref{fig:dE_h4} for $^1$S$_0$ channel.
\label{fig:dE_1S0}
}
\end{figure}

The volume dependences of $\Delta E_L$ 
in the $^3$S$_1$ and $^1$S$_0$ channels
are plotted as a function of $1/L^3$ in Figs.~\ref{fig:dE_3S1}
and ~\ref{fig:dE_1S0}, respectively. 
The numerical values of $\Delta E_L$ on all the spatial volumes
are summarized in Table~\ref{tab:dE_NN}, where the statistical
and systematic errors are given in the first and second parentheses,
respectively.
There is little volume dependence for $\Delta E_L$,  indicating a non-zero negative value in 
the infinite volume and a bound state, rather than the $1/L^3$ dependence expected for a 
scattering state, for the ground state for both channels.

The binding energies in the infinite spatial volume limit 
in Table~\ref{tab:dE_NN} 
are obtained by fitting the data with a 
function including a finite volume effect on the 
two-particle bound state~\cite{Beane:2003da,Sasaki:2006jn},
\begin{equation}
\Delta E_L = -\frac{\gamma^2}{m_N}\left\{
1 + \frac{C_\gamma}{\gamma L} \sum^{\hspace{6mm}\prime}_{\vec{n}}
\frac{\exp(-\gamma L \sqrt{\vec{n}^2})}{\sqrt{\vec{n}^2}}
\right\},
\end{equation}
where $\gamma$ and $C_\gamma$ are free parameters, 
$\vec{n}$ is three-dimensional integer vector, 
and $\sum^\prime_{\vec{n}}$ denotes the summation without $|\vec{n}|=0$.
The binding energy $-\Delta E_\infty$ is determined from
\begin{equation}
-\Delta E_\infty = \frac{\gamma^2}{m_N},
\end{equation}
where we assume
\begin{equation}
2\sqrt{m_N^2 - \gamma^2} - 2 m_N \approx -\frac{\gamma^2}{m_N}.
\end{equation}
The systematic error is estimated from the variation 
of the fit results choosing different fit ranges
in the determination of $\Delta E_L$ and also using constant and linear fits 
as an alternative fit forms.
We obtain the binding energies $-\Delta E_\infty$=11.5(1.1)(0.6) MeV 
and 7.4(1.3)(0.6) MeV for the $^3$S$_1$ and $^1$S$_0$ channels, respectively.
The result for the $^3$S$_1$ channel is roughly five times larger than
the experimental value, 2.22 MeV. 
Our finding of a bound state in the $^1$S$_0$ channel contradicts the experimental observation.
These features are consistent with our quenched results with a heavy
quark mass corresponding to $m_\pi=0.80$ GeV~\cite{Yamazaki:2011nd}.

\subsubsection{Comparison with previous studies}

\begin{figure}[!t]
\includegraphics*[angle=0,width=0.6\textwidth]{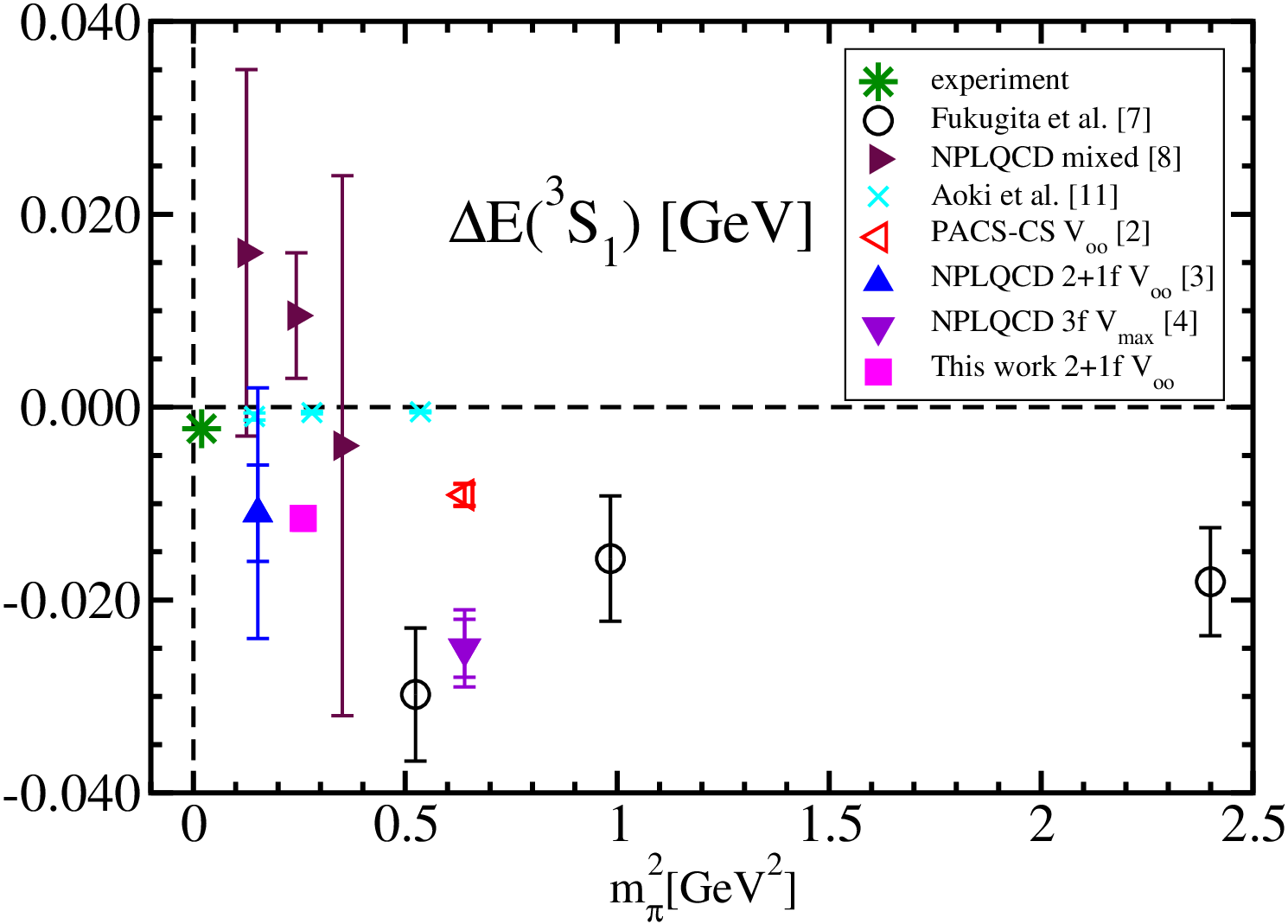}
\caption{
$m_\pi^2$ dependence of $\Delta E_\infty$ for $^3$S$_1$ channel.
Closed(open and cross) symbol denote the 2+1/3 flavor(quenched) result. 
The results of Refs.~\cite{Yamazaki:2011nd,Beane:2011iw} 
and this work are extrapolated values in the infinite volume limit.
Experimental result (star) is also presented for comparison.
\label{fig:mdep_3S1}
}
\end{figure}

\begin{figure}[!t]
\includegraphics*[angle=0,width=0.6\textwidth]{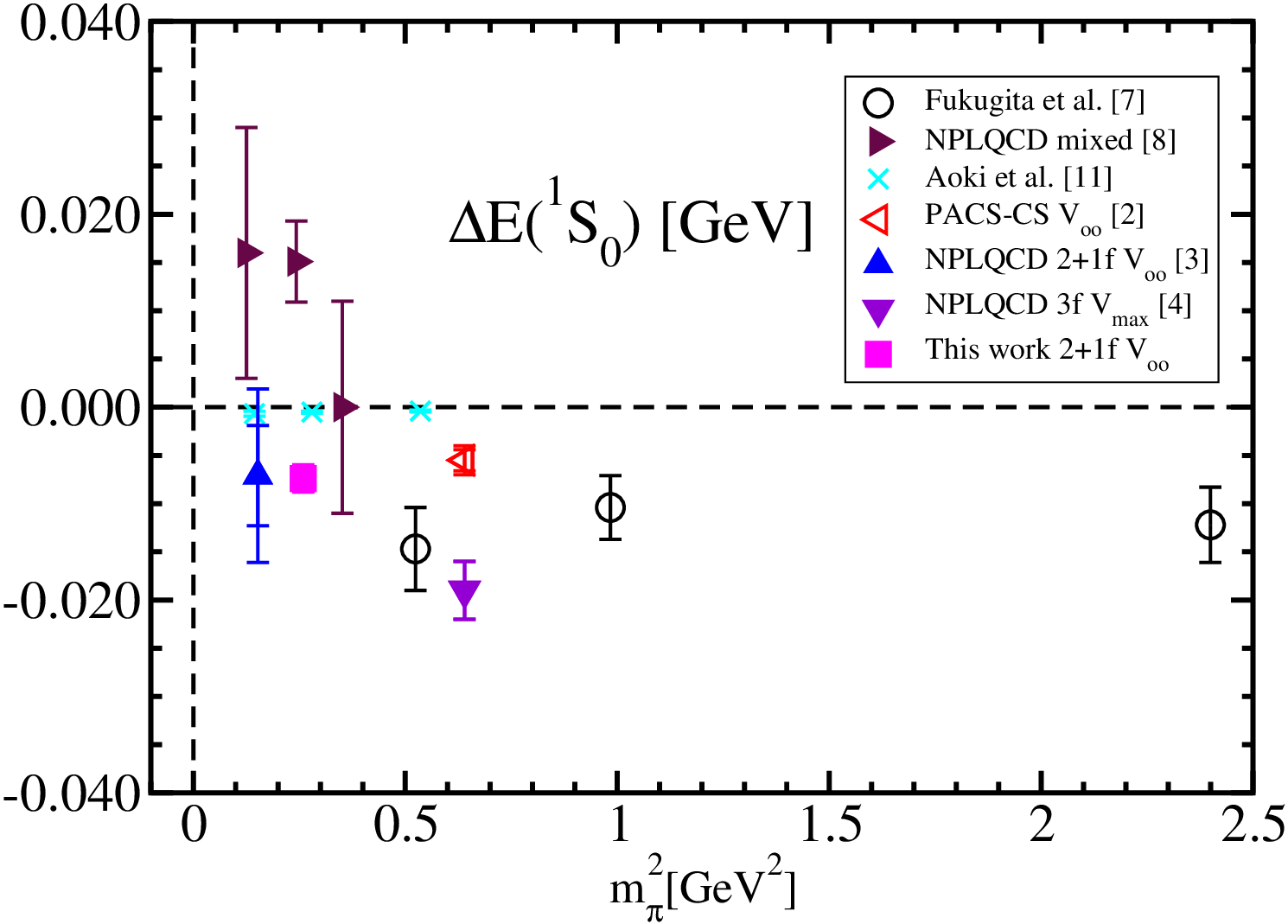}
\caption{
Same as Fig.~\ref{fig:mdep_3S1} for $^1$S$_0$ channel.
\label{fig:mdep_1S0}
}
\end{figure}

A number of studies have been performed for the two-nucleon channel 
after the first work of Ref.~\cite{Fukugita:1994ve}. 
It is therefore instructive to summarize 
the results and make a comparison with each other.
Table~\ref{tab:dE_comp} tabulates, in chronological order,  the results for $-\Delta E_L$ for the
$^3$S$_1$ and $^1$S$_0$ channels together with the pion mass 
$m_\pi$ and the spatial extent $L$ in physical units.
The numbers are plotted in Figs.~\ref{fig:mdep_3S1} and ~\ref{fig:mdep_1S0} 
for the $^3$S$_1$ and $^1$S$_0$ channels, respectively, 
as a function of $m_\pi^2$. 

The early studies in Refs.~\cite{Fukugita:1994ve,Beane:2006mx,Aoki:2009ji} employed 
a single volume, and we do not observe a common feature or trend among them. 
The positive values for 
$-\Delta E_L$ in Ref.~\cite{Beane:2006mx} means repulsive interaction for 
both channels, which is not seen in other studies. 
The results for $-\Delta E_L$ in Ref.~\cite{Aoki:2009ji} 
is an order of magnitude smaller compared to other groups,
probably due to significant contamination from excited states.
 
The four recent studies~\cite{Yamazaki:2011nd,Beane:2011iw,Beane:2012vq} 
have made a systematic investigation of the spatial volume dependence. 
Our quenched and 2+1 flavor results show qualitatively the same feature that 
the binding energy for the $^3$S$_1$ channel is much larger than the 
experimental value and the bound state is observed in the $^1$S$_0$ channel. 
The 2+1 flavor results from Ref.~\cite{Beane:2011iw,Beane:2012vq} 
at $m_\pi = 0.39$ GeV give non-zero negative values
for $\Delta E_L$ in both channels on the $\le$(3.9 fm)$^3$ box, which 
are consistent with our results as shown in Table~\ref{tab:dE_comp}.
Unfortunately, the extrapolation to the infinite spatial volume limit introduces 
large errors so that $\Delta E_\infty$ becomes consistent with zero within the error bars.
The most recent study~\cite{Beane:2012vq} worked  at a heavier quark mass of 
$m_\pi=0.81$ GeV in 3-flavor QCD, and found large values for the binding energies:
25(3)(2) MeV for the $^3$S$_1$ channel 
and 19(3)(1) MeV for the $^1$S$_0$ channel~\cite{Beane:2012vq}.
While all recent studies are consistent with a bound ground state for both $^3$S$_1$ and $^1$S$_0$ channels when quark masses are heavy, quantitative details still need to be clarified.

\section{Conclusion and discussion}
\label{sec:summary}

We have calculated the binding energies for the helium nuclei, the
deuteron and the dineutron in 2+1 flavor QCD with $m_\pi=0.51$ GeV
and $m_N=1.32$ GeV. The bound states are distinguished from the attractive 
scattering states by investigating the spatial volume dependence of 
the energy shift $\Delta E_L$. In the infinite spatial volume limit 
we obtain 
\begin{equation}
-\Delta E_{\infty} = \left\{
\begin{array}{ccl}
43(12)(8) & \mathrm{MeV} & \mathrm{for}\ ^4\mathrm{He},\\
20.3(4.0)(2.0) & \mathrm{MeV} & \mathrm{for}\ ^3\mathrm{He},\\
11.5(1.1)(0.6) & \mathrm{MeV} & \mathrm{for}\ ^3\mathrm{S}_1,\\
7.4(1.3)(0.6) & \mathrm{MeV} & \mathrm{for}\ ^1\mathrm{S}_0.\\
\end{array}
\right.
\end{equation}
While the binding energy for the 
$^4$He nucleus is comparable with the experimental value, those for the 
$^3$He nucleus and the deuteron are much larger than the experimental ones.
Furthermore we detect the bound state in the $^1$S$_0$ channel 
as in the previous study with quenched QCD, which is not observed in nature. 
These findings and the enhanced binding energies at $m_\pi=0.81$ GeV in 3-flavor QCD~\cite{Beane:2012vq} tell us that a next step of primary importance
is to reduce the up-down quark mass toward the physical values.
A possible scenario in the two-nucleon channels is as follows.
The binding energy in both channels
diminishes monotonically as the up-down quark mass decreases.
At some point of the up-down quark mass the binding energy in 
the $^1$S$_0$ channel vanishes and the bound state evaporates into 
the attractive scattering state, while the binding energy in the 
$^3$S$_1$ channel remains finite up to the physical point. 
This is a dynamical question on the strong interaction,  and only lattice QCD
could answer it.
\section*{Acknowledgments}
Numerical calculations for the present work have been carried out
on the HA8000 cluster system at Information Technology Center
of the University of Tokyo, on the PACS-CS computer 
under the ``Interdisciplinary Computational Science Program'' of 
Center for Computational Sciences in University of Tsukuba, 
on the T2K-Tsukuba cluster system and HA-PACS system at University of Tsukuba,
and on K computer at RIKEN Advanced Institute for Computational Science.
We thank our colleagues in the PACS-CS Collaboration for helpful
discussions and providing us the code used in this work.
This work is supported in part by Grants-in-Aid for Scientific Research
from the Ministry of Education, Culture, Sports, Science and Technology 
(Nos. 18104005, 18540250, 22244018) and 
Grants-in-Aid of the Japanese Ministry for Scientific Research on Innovative 
Areas (Nos. 20105002, 21105501, 23105708).

\bibliography{paperHe-NN}

\clearpage
%
%
% Tables
%
%
\begin{table}[!t]
\caption{
Simulation parameters for gauge configuration generation at 
($\kappa_{ud}$, $\kappa_s$)=(0.1373316,0.1367526).
The definition of parameters is same as in Ref.~\cite{Aoki:2008sm}.
\label{tab:conf_gene}
}
\begin{ruledtabular}
\begin{tabular}{ccccc}
$L^3 \times T$& $32^3\times 48$ & $40^3\times 48$ & $48^3\times 48$ &
 $64^3\times 64$\\
\# run         & 1 & 1 & 1 & 2 \\
$(N_0,N_1,N_2)$ & (2,2,10) & (2,2,15) & (2,2,16) & (2,2,18) \\
Block size   
 & $8^3\times 6$ & $10^3\times 6$ & $12^2\times 6^2$ & $8^3\times 4$\\
$N_{\rm poly}$ & 260 & 320 & 320 & 340 \\
MD time       & 4000 & 2000 & 2000 & (1090,810) \\
$P_{acc}$(HMC)& 0.840 & 0.925 & 0.916 & (0.880,0.867) \\
$P_{acc}$(GMP)& 0.957 & 0.969 & 0.963 & (0.978,0.974) \\
\end{tabular}
\end{ruledtabular}
\end{table}
\begin{table}[!t]
\caption{
Number of configurations, 
separation of trajectories between each measurement,
bin size in jackknife analysis,
number of measurements on each configuration,
exponential smearing parameter set ($A,B$) in Eq.~(\ref{eq:smear}),
pion mass $m_\pi$ and nucleon mass $m_N$ are summarized
for each lattice size.
\label{tab:conf_meas}
}
\begin{ruledtabular}
\begin{tabular}{ccccccccc}
$L$ & $T$ & \# config. & $\tau_{\mathrm{sep}}$ 
& bin size & \# meas. & $(A,B)$
& $m_\pi$ [GeV] & $m_N$ [GeV]\\
\hline
32 & 48 & 200 & 20 & 10 & 192 & (1.0,0.18) & 0.5109(16) & 1.318(4) \\
40 & 48 & 200 & 10 & 20 & 192 & (0.8,0.22) & 0.5095(8)  & 1.314(4) \\
48 & 48 & 200 & 10 & 20 & 192 & (0.8,0.23) & 0.5117(9)  & 1.320(3) \\
64 & 64 & 190 & 10 & 19 & 256 & (0.8,0.23) & 0.5119(4)  & 1.318(2) \\
\end{tabular}
\end{ruledtabular}
\end{table}
\begin{table}[!t]
\caption{
\label{tab:dE_He}
Energy shift $-\Delta E_L$ in physical units 
and fit range for $^4$He and $^3$He channels 
on each spatial volume.
Extrapolated results in the infinite spatial volume limit 
are also presented. The first and second errors are
statistical and systematic, respectively.
}
\begin{ruledtabular}
\begin{tabular}{ccccc}
& \multicolumn{2}{c}{$^4$He}&
\multicolumn{2}{c}{$^3$He}\\
$L$  & $-\Delta E_L$ [MeV] & fit range & $-\Delta E_L$ [MeV] & fit range \\
\hline
32 & 47(24)(5)  & 10--14 & 23.2(7.6)(1.4) & 10--14 \\
40 & 30(15)(23) &  9--13 & 20.2(6.9)(2.8) &  9--14 \\
48 & 39(20)(27) & 10--14 & 25.5(5.3)(1.7) & 10--14 \\
64 & 46(11)(8)  & 10--14 & 19.5(3.7)(1.2) &  9--14 \\
$\infty$ & 43(12)(8) & --- &
20.3(4.0)(2.0) & --- \\
\end{tabular}
\end{ruledtabular}
\end{table}
\begin{table}[!t]
\caption{
\label{tab:dE_NN}
Same as Table~\ref{tab:dE_He} for $^3$S$_1$ and $^1$S$_0$ channels.
}
\begin{ruledtabular}
\begin{tabular}{ccccc}
& \multicolumn{2}{c}{$^3$S$_1$}&
\multicolumn{2}{c}{$^1$S$_0$}\\
$L$  & $-\Delta E_L$ [MeV] & fit range & $-\Delta E_L$ [MeV] & fit range \\
\hline
32 & 12.4(2.1)(0.5) &  9--14 & 6.2(2.4)(0.5) & 10--14 \\
40 & 12.2(1.9)(0.6) & 10--15 & 8.2(4.0)(1.5) & 11--15 \\
48 & 11.1(1.7)(0.3) & 10--14 & 7.3(1.7)(0.5) & 10--14 \\
64 & 11.7(1.2)(0.5) &  9--14 & 7.2(1.4)(0.3) & 10--14 \\
$\infty$ & 11.5(1.1)(0.6) & --- &
7.4(1.3)(0.6) & --- \\
\end{tabular}
\end{ruledtabular}
\end{table}
\begin{table}[!t]
\caption{
\label{tab:dE_comp}
Energy shift $-\Delta E_L$ in physical units 
for $^3$S$_1$ and $^1$S$_0$ channels 
together with the previous works. 
The values marked by $*$ are estimated from
the scattering length in Ref.~\cite{Beane:2006mx} employing
the leading term of finite volume formula
in the $1/L$ expansion~\cite{Luscher:1986pf} with 
the nucleon mass obtained from
the same ensemble in Ref.~\cite{Bratt:2010jn}.
The values for Ref.~\cite{Yamazaki:2011nd} are taken from the results 
with the ${\cal O}_1$ interpolating operator.
}
\begin{ruledtabular}
\begin{tabular}{ccccccc}
Ref. & quark action & \# flavor & $m_\pi$ [GeV] & $L$ [fm] & \multicolumn{2}{c}{$-\Delta E_L$ [MeV]} \\
 &  & & & & $^3$S$_1$ & $^1$S$_0$ \\\hline
\cite{Fukugita:1994ve} & Wilson & 0 & 0.72 & 2.7 & 29.8(6.9) & 14.7(4.3) \\
                       && 0 & 0.99 & 2.7 & 15.7(6.5) & 10.7(4.3) \\
                       && 0 & 1.55 & 2.7 & 18.1(5.6) & 12.2(3.9) \\
\cite{Beane:2006mx}  & Mixed (DW on Asqtad) & 2+1 & 0.35 & 2.5 & $-$16(19)$^*$ & $-$16(13)$^*$ \\
                     && 2+1 & 0.49 & 2.5 & $-$9.5(6.5)$^*$ & $-$15.1(4.2)$^*$ \\
                     && 2+1 & 0.59 & 2.5 & 0.4(2.8)$^*$ & 0.0(1.1)$^*$ \\
\cite{Aoki:2009ji}    &Wilson& 0 & 0.38 & 4.4 & 0.97(37) & 0.68(26) \\
                       && 0 & 0.53 & 4.4 & 0.56(11) & 0.509(94) \\
                       && 0 & 0.73 & 4.4 & 0.480(97) & 0.400(83) \\
\cite{Yamazaki:2011nd} &Wilson-clover& 0 & 0.80 & 3.1 & 10.2(2.2)(1.6) & 6.1(2.3)(2.2) \\
                       && 0 & 0.80 & 6.1 &  9.6(2.6)(0.9) & 5.2(2.6)(0.8) \\
                       && 0 & 0.80 & 12.3 & 7.8(2.1)(0.4) & 4.6(2.0)(1.1) \\
                       && 0 & 0.80 & $\infty$ & 9.1(1.1)(0.5) & 5.5(1.1)(1.0) \\
\cite{Beane:2009py}  & Aniso. Wislon-clover& 2+1 & 0.39 & 2.4 & 1.6(2.6)(4.3) & 3.9(1.7)(2.6) \\
\cite{Beane:2011iw}  & Aniso. Wilson-clover & 2+1 & 0.39 & 3.0 & 22.3(2.3)(5.4) & 10.4(2.6)(3.1) \\
                     && 2+1 & 0.39 & 3.9 & 14.9(2.3)(5.8) & 8.3(2.2)(3.3) \\
                     && 2+1 & 0.39 & $\infty$ & 11(5)(12) & 7.1(5.2)(7.3) \\
\cite{Beane:2012vq}  & Stout Wilson-clover & 2+1 & 0.81 & 3.4 & 25(3)(2) & 16(3)(1) \\
                     && 2+1 & 0.81 & 4.5 & 21(3)(1) & 11(2)(1) \\
                     && 2+1 & 0.81 & 6.7 & 25(3)(2) & 19(3)(1) \\
This work            & Wilson-clover& 2+1 & 0.51 & 2.9 & 12.4(2.1)(0.5) & 6.2(2.4)(0.5)\\
                     && 2+1 & 0.51 & 3.6 & 12.2(1.9)(0.6) & 8.2(4.0)(1.5)\\
                     && 2+1 & 0.51 & 4.3 & 11.1(1.7)(0.3) & 7.3(1.7)(0.5)\\
                     && 2+1 & 0.51 & 5.8 & 11.7(1.2)(0.5) & 7.2(1.4)(0.3)\\
                     && 2+1 & 0.51 & $\infty$ & 11.5(1.1)(0.6) & 7.4(1.3)(0.6)\\
\end{tabular}
\end{ruledtabular}
\end{table}

\clearpage
%
%
% Figures
%
%

\end{document}